\documentclass[12pt,a4paper]{article}
\usepackage{lineno}
\ifx\pdfoutput\undefined
\usepackage{graphicx}
\else
\usepackage[pdftex]{graphicx}
\usepackage{epstopdf}
\fi

\usepackage{xspace}
\usepackage{color}
\usepackage{colortbl}

\usepackage{amsmath}

\usepackage{ifthen}
\newboolean{pdflatex}
\setboolean{pdflatex}{false} 
\newboolean{articletitles}
\setboolean{articletitles}{true} 

\newboolean{uprightparticles}
\setboolean{uprightparticles}{false} 
\usepackage{amssymb}
\usepackage{amsfonts}
\usepackage{upgreek}

\usepackage{hyperref}
\usepackage[all]{hypcap} 

\def\lhcb {LHCb\xspace}
\def\ux85 {UX85\xspace}

\def\lhc {LHC\xspace}

\def\belle  {BELLE\xspace}

\ifthenelse{\boolean{uprightparticles}}%
{

 \def\Ppi         {\ensuremath{\uppi}\xspace}

 \def\PDelta      {\ensuremath{\Delta}\xspace}                 
 \def\PXi      {\ensuremath{\Xi}\xspace}                 
 \def\PLambda      {\ensuremath{\Lambda}\xspace}                 
 \def\PSigma      {\ensuremath{\Sigma}\xspace}                 
 \def\POmega      {\ensuremath{\Omega}\xspace}                 
 \def\PUpsilon      {\ensuremath{\Upsilon}\xspace}                 
 
 \def\PB      {\ensuremath{\mathrm{B}}\xspace}                 
                  
 \def\PD      {\ensuremath{\mathrm{D}}\xspace}

 \def\PK      {\ensuremath{\mathrm{K}}\xspace}

 \def\Pb      {\ensuremath{\mathrm{b}}\xspace}                 
 \def\Pc      {\ensuremath{\mathrm{c}}\xspace}

 \def\Pi      {\ensuremath{\mathrm{i}}\xspace}

 \def\Pp      {\ensuremath{\mathrm{p}}\xspace}

}
{

 \def\Ppi         {\ensuremath{\pi}\xspace}

 \mathchardef\PDelta="7101
 \mathchardef\PXi="7104
 \mathchardef\PLambda="7103
 \mathchardef\PSigma="7106
 \mathchardef\POmega="710A
 \mathchardef\PUpsilon="7107
                  
 \def\PB      {\ensuremath{B}\xspace}                 
                  
 \def\PD      {\ensuremath{D}\xspace}

 \def\PK      {\ensuremath{K}\xspace}

 \def\Pb      {\ensuremath{b}\xspace}                 
 \def\Pc      {\ensuremath{c}\xspace}

 \def\Pi      {\ensuremath{i}\xspace}

 \def\Pp      {\ensuremath{p}\xspace}

}

\def\cquark    {\ensuremath{\Pc}\xspace}

\def\bquark    {\ensuremath{\Pb}\xspace}

\def\pion  {\ensuremath{\Ppi}\xspace}

\def\pip   {\ensuremath{\pion^+}\xspace}
\def\pim   {\ensuremath{\pion^-}\xspace}

\def\kaon  {\ensuremath{\PK}\xspace}
  \def\Kbar  {\kern 0.2em\overline{\kern -0.2em \PK}{}\xspace}

\def\Kz    {\ensuremath{\kaon^0}\xspace}
\def\Kzb   {\ensuremath{\Kbar^0}\xspace}
\def\KzKzb {\ensuremath{\Kz \kern -0.16em \Kzb}\xspace}
\def\Kp    {\ensuremath{\kaon^+}\xspace}
\def\Km    {\ensuremath{\kaon^-}\xspace}

\def\KpKm  {\ensuremath{\Kp \kern -0.16em \Km}\xspace}

  \def\Dbar    {\kern 0.2em\overline{\kern -0.2em \PD}{}\xspace}
\def\D       {\ensuremath{\PD}\xspace}

\def\Dz      {\ensuremath{\D^0}\xspace}
\def\Dzb     {\ensuremath{\Dbar^0}\xspace}
\def\DzDzb   {\ensuremath{\Dz {\kern -0.16em \Dzb}}\xspace}
\def\Dp      {\ensuremath{\D^+}\xspace}
\def\Dm      {\ensuremath{\D^-}\xspace}

\def\DpDm    {\ensuremath{\Dp {\kern -0.16em \Dm}}\xspace}

\def\Dstarp  {\ensuremath{\D^{*+}}\xspace}

\def\Dsm     {\ensuremath{\D^-_s}\xspace}

\def\Dssm    {\ensuremath{\D^{*-}_s}\xspace}

\def\B       {\ensuremath{\PB}\xspace}
  \def\Bbar    {\kern 0.18em\overline{\kern -0.18em \PB}{}\xspace}

\def\Bz      {\ensuremath{\B^0}\xspace}

\def\Bs      {\ensuremath{\B^0_s}\xspace}

  \def\Y#1S{\ensuremath{\PUpsilon{(#1S)}}\xspace}% no space before {...}!

\def\proton      {\ensuremath{\Pp}\xspace}
\def\antiproton  {\ensuremath{\overline \proton}\xspace}

\def\L {\ensuremath{\PLambda}\xspace}
\def\Lbar{\ensuremath{\overline \L}\xspace}

\def\Lb      {\ensuremath{\L^0_b}\xspace}
\def\Lbbar   {\ensuremath{\Lbar^0_b}\xspace}

\def\Lcbar   {\ensuremath{\Lbar_c^-}\xspace}

\def\BF         {{\ensuremath{\cal B}\xspace}}

\def\BR         {\BF}
\newcommand{\decay}[2]{\ensuremath{#1\!\to #2}\xspace}         % {\Pa}{\Pb \Pc}

\def\to                 {\ensuremath{\rightarrow}\xspace}

\def\CP                {\ensuremath{C\!P}\xspace}

\def\AT#1     {\ensuremath{A_T^{#1}}\xspace}           % 2

\def\C#1      {\ensuremath{\mathcal{C}_{#1}}\xspace}                       % 9
\def\Cp#1     {\ensuremath{\mathcal{C}_{#1}^{'}}\xspace}                    % 7
\def\Ceff#1   {\ensuremath{\mathcal{C}_{#1}^{\mathrm{(eff)}}}\xspace}        % 9  
\def\Cpeff#1  {\ensuremath{\mathcal{C}_{#1}^{'\mathrm{(eff)}}}\xspace}       % 7
\def\Ope#1    {\ensuremath{\mathcal{O}_{#1}}\xspace}                       % 2
\def\Opep#1   {\ensuremath{\mathcal{O}_{#1}^{'}}\xspace}                    % 7

\newcommand{\tev}{\ensuremath{\mathrm{\,Te\kern -0.1em V}}\xspace}
\newcommand{\gev}{\ensuremath{\mathrm{\,Ge\kern -0.1em V}}\xspace}
\newcommand{\mev}{\ensuremath{\mathrm{\,Me\kern -0.1em V}}\xspace}
\newcommand{\kev}{\ensuremath{\mathrm{\,ke\kern -0.1em V}}\xspace}
\newcommand{\ev}{\ensuremath{\mathrm{\,e\kern -0.1em V}}\xspace}
\newcommand{\gevc}{\ensuremath{{\mathrm{\,Ge\kern -0.1em V\!/}c}}\xspace}
\newcommand{\mevc}{\ensuremath{{\mathrm{\,Me\kern -0.1em V\!/}c}}\xspace}
\newcommand{\gevcc}{\ensuremath{{\mathrm{\,Ge\kern -0.1em V\!/}c^2}}\xspace}
\newcommand{\gevgevcccc}{\ensuremath{{\mathrm{\,Ge\kern -0.1em V^2\!/}c^4}}\xspace}
\newcommand{\mevcc}{\ensuremath{{\mathrm{\,Me\kern -0.1em V\!/}c^2}}\xspace}

\def\mum  {\ensuremath{\,\upmu\rm m}\xspace}

\newcommand{\chisq}{\ensuremath{\chi^2}\xspace}

\def\gsim{{~\raise.15em\hbox{$>$}\kern-.85em
          \lower.35em\hbox{$\sim$}~}\xspace}
\def\lsim{{~\raise.15em\hbox{$<$}\kern-.85em
          \lower.35em\hbox{$\sim$}~}\xspace}

\def\evtgen     {\mbox{\textsc{EvtGen}}\xspace}
\def\pythia     {\mbox{\textsc{Pythia}}\xspace}

\def\geant      {\mbox{\textsc{Geant4}}\xspace}

\def\tell1  {TELL1\xspace}
\def\ukl1   {UKL1\xspace}

\newcommand{\eg}{\mbox{\itshape e.g.}\xspace}

\newcommand{\BdDp} {\decay{\Bz}{\Dm \pip}}
\newcommand{\BdDsp} {\decay{\Bz}{\Dsm \pip}}
\newcommand{\BdDK} {\decay{\Bz}{\Dm \Kp}}
\newcommand{\BsDp} {\decay{\Bs}{\Dsm \pip}}

\newcommand{\BsDK} {\decay{\Bs}{D_s^\mp K^\pm}}
\newcommand{\BdDsK} {\decay{\Bz}{\Dsm \Kp}}

\newcommand{\BDK}  {\decay{\Bz}{\Dm \Kp}}

\newcommand{\LbLcp}    {\decay{\Lbbar}{\Lcbar \pip}}
\newcommand{\LbDsp}    {\decay{\Lb}{\Dsm p}}
\newcommand{\LbDsstp}    {\decay{\Lb}{\Dssm p}}

\newcommand{\BsDstarp} {\decay{\Bs}{\Dssm \pip}}
\newcommand{\BsDstarK} {\decay{\Bs}{\Dssm \Kp}}
\newcommand{\BsDrho}   {\decay{\Bs}{\Dsm \rho^{+}}}
\newcommand{\BsDkst}   {\decay{\Bs}{\Dsm K^{*+}}}
\newcommand{\BsDstrho}   {\decay{\Bs}{\Dssm \rho^{+}}}
\newcommand{\BsDstkst}   {\decay{\Bs}{\Dssm K^{*+}}}

\newcommand{\fsfd}{\ensuremath{\frac{f_s}{f_d}}}
\newcommand{\fsfdt}{\ensuremath{f_s/f_d}\xspace}

\usepackage{cite}
\usepackage{mciteplus}
\begin{document}
\newcommand\TVA{\rule{0pt}{3.6ex}}
\newcommand\BVA{\rule[-2.2ex]{0pt}{0pt}}

\renewcommand{\thefootnote}{\fnsymbol{footnote}}
\setcounter{footnote}{1}
% $Id: title-LHCb-PAPER.tex 14593 2012-01-27 14:43:32Z uegede $
% ===============================================================================
% Purpose: LHCb-PAPER journal paper title page template
% Author: 
% Created on: 2010-09-25
% ===============================================================================

%%%%%%%%%%%%%%%%%%%%%%%%%
%%%%%  TITLE PAGE  %%%%%%
%%%%%%%%%%%%%%%%%%%%%%%%%
\begin{titlepage}
\pagenumbering{roman}

% Header ---------------------------------------------------
\vspace*{-1.5cm}
\centerline{\large EUROPEAN ORGANIZATION FOR NUCLEAR RESEARCH (CERN)}
\vspace*{1.5cm}
\hspace*{-0.5cm}
\begin{tabular*}{\linewidth}{lc@{\extracolsep{\fill}}r}
\ifthenelse{\boolean{pdflatex}}% Logo format choice
{\vspace*{-2.7cm}\mbox{\!\!\!\includegraphics[width=.14\textwidth]{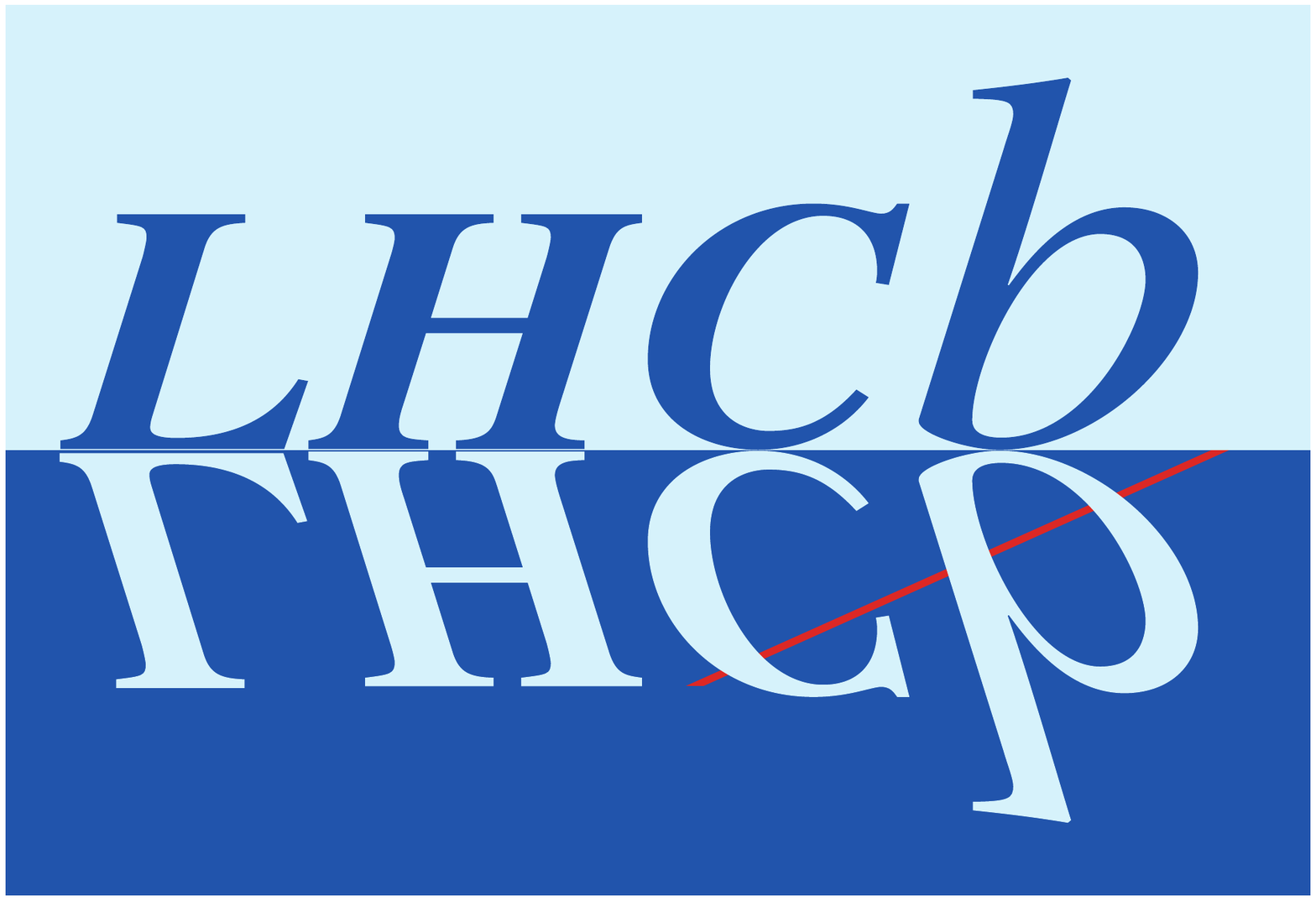}} & &}
{\vspace*{-1.2cm}\mbox{\!\!\!\includegraphics[width=.12\textwidth]{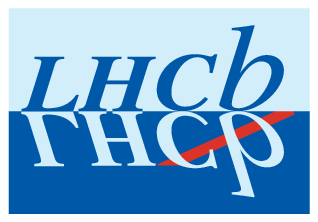}} & &}
\\
 & & CERN-PH-EP-2012-091 \\  
 & & LHCb-PAPER-2011-022 \\  
 & & 5 April 2012 \\ 
 & & \\
\end{tabular*}

\vspace*{1.0cm}

{\bf\boldmath\huge
\begin{center}
Measurements of the branching fractions
of the decays \BsDK and \BsDp
\end{center}
}

\vspace*{1.0cm}

\begin{center}
LHCb collaboration
\footnote{Authors are listed on the following pages.}
\end{center}

\vspace{\fill}

\begin{abstract}
  \noindent
The decay mode \BsDK allows for one of the theoretically 
cleanest measurements of the CKM angle 
$\gamma$ through the study of time-dependent \CP violation.
This paper reports a measurement of its branching fraction relative to the Cabibbo-favoured mode 
\BsDp based on a data sample of 0.37~fb$^{-1}$ proton-proton collisions at
$\sqrt{s} = 7$ TeV collected in 2011 with the \lhcb detector. 
In addition, the ratio of \B meson production
fractions \fsfdt, determined from semileptonic decays,
together with the known branching fraction of the control channel \BdDp, is used to perform an absolute measurement of the branching fractions:

\begin{center}
$\BR\left(\BsDp \right) \;= (2.95 \pm 0.05 \pm 0.17^{\,+\,0.18}_{\,-\,0.22}) \times 10^{-3}\,$, 
\end{center}

\begin{center}
$\BR\left(\BsDK \right) = (1.90 \pm 0.12 \pm 0.13^{\,+\,0.12}_{\,-\,0.14}) \times 10^{-4}\,$,
\end{center}
where the first uncertainty is statistical, the second the experimental systematic uncertainty, and the
third the uncertainty due to \fsfdt. 
\end{abstract}

\vspace*{0.5cm}

\begin{center}
  Submitted to JHEP
\end{center}

\vspace{\fill}

\end{titlepage}

%%%%%%%%%%%%%%%%%%%%%%%%%%%%%%%%
%%%%%  EOD OF TITLE PAGE  %%%%%%
%%%%%%%%%%%%%%%%%%%%%%%%%%%%%%%%

\setcounter{page}{2}
\mbox{~}
%%%%%%%%%%%%%%%%%%%%%%%%%%%%%%%%%%%%%%%%%%
\centerline{\large\bf LHCb collaboration}
\begin{flushleft}
\small
R.~Aaij$^{38}$, 
C.~Abellan~Beteta$^{33,n}$, 
B.~Adeva$^{34}$, 
M.~Adinolfi$^{43}$, 
C.~Adrover$^{6}$, 
A.~Affolder$^{49}$, 
Z.~Ajaltouni$^{5}$, 
J.~Albrecht$^{35}$, 
F.~Alessio$^{35}$, 
M.~Alexander$^{48}$
% ...

{\footnotesize \it
$ ^{1}$Centro Brasileiro de Pesquisas F\'{i}sicas (CBPF), Rio de Janeiro, Brazil\\
% ...
$ ^{5}$Clermont Universit\'{e}, Universit\'{e} Blaise Pascal, CNRS/IN2P3, LPC, Clermont-Ferrand, France\\
$ ^{6}$CPPM, Aix-Marseille Universit\'{e}, CNRS/IN2P3, Marseille, France\\
% ...
$ ^{33}$Universitat de Barcelona, Barcelona, Spain\\
$ ^{34}$Universidad de Santiago de Compostela, Santiago de Compostela, Spain\\
$ ^{35}$European Organization for Nuclear Research (CERN), Geneva, Switzerland\\
% ...
$ ^{38}$Nikhef National Institute for Subatomic Physics, Amsterdam, The Netherlands\\
% ...
$ ^{43}$H.H. Wills Physics Laboratory, University of Bristol, Bristol, United Kingdom\\
% ...
$ ^{48}$School of Physics and Astronomy, University of Glasgow, Glasgow, United Kingdom\\
$ ^{49}$Oliver Lodge Laboratory, University of Liverpool, Liverpool, United Kingdom\\
% ...
$ ^{n}$LIFAELS, La Salle, Universitat Ramon Llull, Barcelona, Spain\\
% ...
}
\end{flushleft}
%%%%%%%%%%%%%%%%%%%%%%%%%%%%%%%%%%%%%%%%%%

\cleardoublepage

\renewcommand{\thefootnote}{\arabic{footnote}}
\setcounter{footnote}{0}
\pagestyle{plain} 
\setcounter{page}{1}
\pagenumbering{arabic}
\section{Introduction}
\label{sec:Introduction}

Unlike the flavour-specific decay \BsDp, the Cabibbo-suppressed decay \BsDK proceeds 
through two different tree-level amplitudes of similar strength: a $\bar{b} \to \bar{c} u \bar{s}$ 
transition leading to $B^0_s \to D^-_s K^+$ and a $\bar{b} \to   \bar{u} c \bar{s}$ transition 
leading to $B^0_s \to   D^+_sK^-$. These two decay amplitudes can have a large \CP-violating 
interference via $B^0_s-\bar{B}^0_s$ mixing, allowing the determination of the CKM angle $\gamma$ 
with negligible theoretical uncertainties through the measurement of tagged and untagged time-dependent 
decay rates to both the $D^-_s K^+$ and $D^+_s K^-$ final states \cite{Fleischer}. 
Although the \BsDK decay mode has been observed by the CDF~\cite{Aaltonen:2008rw} and \belle~\cite{BelleBsDspi} collaborations,
only the \lhcb experiment 
has both the necessary decay time resolution and access to large enough signal yields to perform the time-dependent \CP measurement.
In this analysis, the \BsDK branching fraction is determined relative to \BsDp, and the absolute \BsDp branching fraction
is determined using the known branching fraction of \BdDp and 
the production fraction ratio \fsfdt \cite{Aaij:2011jp}. 
The two measurements are then combined to obtain the absolute branching
fraction of the decay \BsDK. Charge conjugate modes are implied throughout. 
Our notation \BdDp, which matches that of Ref.~\cite{PDG}, encompasses both the Cabibbo-favoured 
$B^0 \to D^-\pi^+$ mode and the doubly-Cabibbo-suppressed $B^0 \to D^+\pi^-$ mode.

The \lhcb detector~\cite{DetectPaper} is a single-arm forward
spectrometer covering the pseudo-rapidity range $2<\eta <5$, designed
for studing particles containing \bquark or \cquark quarks. 
The 
detector includes a high-precision tracking system consisting of a
silicon-strip vertex detector surrounding the $pp$ interaction region,
a large-area silicon-strip detector located upstream of a dipole
magnet with a bending power of about $4{\rm\,Tm}$, and three stations
of silicon-strip detectors and straw drift tubes placed
downstream. The combined tracking system has a momentum resolution
$\Delta p/p$ that varies from 0.4\% at 5\gevc to 0.6\% at 100\gevc,
an impact parameter resolution of 20\mum for tracks with high
transverse momentum, and a decay time resolution of 50~fs. Charged hadrons are identified using two 
ring-imaging Cherenkov detectors. Photon, electron and hadron
candidates are identified by a calorimeter system consisting of
scintillating-pad and pre-shower detectors, an electromagnetic
calorimeter, and a hadronic calorimeter. Muons are identified by a muon
system composed of alternating layers of iron and multiwire
proportional chambers.

The \lhcb trigger consists of a hardware stage, based
on information from the calorimeter and muon systems, followed by a
software stage which applies a full event reconstruction.
Two categories of events are recognised based on the hardware trigger decision. The first category are 
events triggered by tracks from signal decays which have an associated cluster in the calorimeters,
and the second category are events triggered independently of the signal decay particles.
Events which do not fall into either of these two categories are not used in the subsequent analysis.
The second, software, trigger stage requires a two-, three- or four-track
secondary vertex with a large value of the scalar sum of the transverse momenta ($p_{\rm T}$) 
of the tracks, and a significant
displacement from the primary interaction. At least one of the tracks used to form
this vertex is required to have $p_{\rm T}> 1.7$~GeV$/c$, an impact parameter \chisq $>16$, and a track fit $\chisq$
per degree of freedom $\chisq/\textrm{ndf}$ $< 2$. 
A multivariate algorithm is used for the identification of the secondary
vertices~\cite{Gligorov:1384380}. Each input variable is binned to minimise
the effect of systematic differences between the trigger behaviour on data and simulated events. 

The samples of simulated events used in this analysis
are based on the \pythia 6.4 generator \cite{Sjostrand:2006za}, with a choice of parameters specifically configured for LHCb \cite{procgenerator}. The \evtgen package \cite{Lange:2001uf}
describes the decay of the \B mesons, and the \geant package \cite{Agostinelli:2002hh} simulates the
detector response. QED radiative corrections are generated with the  {\mbox{\textsc{Photos}}\xspace} package~\cite{Golonka:2005pn}.

The analysis is based on a sample of $pp$ collisions corresponding to an integrated luminosity of 0.37~fb$^{-1}$,  
collected at the \lhc in 2011 at a centre-of-mass energy \mbox{$\sqrt{s} = 7$ TeV}.
The decay modes \BsDp and \BsDK are topologically identical and are selected
using identical geometric and kinematic criteria, thereby minimising efficiency
corrections in the ratio of branching fractions. The decay mode \BdDp has a similar
topology to the other two, differing only in the Dalitz plot structure of the \D decay and
the lifetime of the \D meson. These differences are verified, using simulated events, to alter the selection efficiency
at the level of a few percent, and are taken into account.

\Bs (\Bz) candidates are reconstructed from a \Dsm (\Dm) candidate 
and an additional pion or kaon (the ``bachelor'' particle), with
the \Dsm (\Dm) meson decaying in the $K^+ K^- \pi^-$ ($K^+ \pi^-\pi^-$) mode.
All selection criteria will now be specified for the \Bs decays, and are implied to be
identical for the \Bz decay unless explicitly stated otherwise.
All final-state particles are required to satisfy a track fit $\chi^2 / \textrm{ndf} < 4$ and to
have a high transverse momentum and a large impact parameter $\chi^2$ with
respect to all primary vertices in the event.
In order to remove backgrounds which contain the same final-state particles as the signal decay, and therefore
have the same mass lineshape, but do not proceed through the decay of a charmed meson,
the flight distance \chisq of the \Dsm from the \Bs is required to be larger than $2$.
Only \Dsm and bachelor candidates forming a vertex with a 
$\chi^2 / \textrm{ndf} < 9$ are considered as \Bs candidates.
The same vertex quality criterion is applied to the \Dsm candidates.
The \Bs candidate is further required to point to the primary vertex
imposing $\theta_{\textrm{flight}} < 0.8$ degrees, where $\theta_{\textrm{flight}}$
is the angle between the candidate momentum vector and 
the line between the primary vertex and the \Bs vertex.
The \Bs candidates are also required to have a $\chi^2$ of their impact parameter
with respect to the primary vertex less than 16.

Further suppression of combinatorial backgrounds is achieved
using a gradient boosted decision tree
technique~\cite{TMVA} identical to the decision tree used in the previously
published determination of \fsfdt~with the hadronic decays \cite{PhysRevLett.107.211801}.
The optimal working point is evaluated directly from a sub-sample of \BsDp events,
corresponding to 10$\%$ of the full dataset used, distributed evenly over the
data taking period and selected using particle identification and trigger requirements.
The chosen figure of merit is the significance of the \BsDK signal, scaled according
to the Cabibbo suppression relative to the \BsDp signal, with respect to the combinatorial background.
The significance exhibits a wide plateau around its maximum, and the optimal working point
is chosen at the point in the plateau which maximizes the signal yield.
Multiple candidates occur in about $2\%$ of the events and in such cases
a single candidate is selected at random.

\section{Particle identification}
\label{sec:PID}
Particle identification (PID) criteria serve two purposes in the
selection of the three signal decays \BdDp, \BsDp and \BsDK.
When applied to the decay products of the \Dsm or \Dm, they suppress misidentified backgrounds
which have the same bachelor particle as the signal mode
under consideration, henceforth the ``cross-feed'' backgrounds. When applied to the bachelor particle
(pion or kaon) they separate the Cabibbo-favoured  
from the Cabibbo-suppressed decay modes. 
All PID criteria are based on the differences in log-likelihood (DLL) between
the kaon, proton, or pion hypotheses. Their efficiencies are obtained
from calibration samples of $\Dstarp\to (\Dz\to K^- \pip) \pip$ and $\L\to p\pim$ signals, which are themselves
selected without any PID requirements. These samples are 
split according to the magnet polarity, binned in momentum and $p_{\rm T}$,
and then reweighted to have the same momentum and $p_{\rm T}$
distributions as the signal decays under study.

The selection of a pure \BdDp sample can be accomplished with minimal PID requirements since all cross-feed backgrounds are
less abundant than the signal. The \LbLcp background is suppressed by requiring
that both pions produced in the \Dm decay satisfy $\textrm{DLL}_{\pi-p}>-10$, and the
\BdDK background is suppressed by requiring that the bachelor pion satisfies 
$\textrm{DLL}_{K-\pi}<0$.

The selection of a pure \BsDp or \BsDK sample requires the suppression of the \BdDp and
\LbLcp backgrounds, whereas the combinatorial background contributes to a lesser extent.
The $D^{-}$ contamination in the $D^{-}_{s}$ data sample is reduced by requiring that the kaon
which has the same charge as the pion in $\Dsm \to K^+ K^- \pim$ satisfies 
$\textrm{DLL}_{K-\pi}>5$. In addition, the other kaon is required to satisfy 
$\textrm{DLL}_{K-\pi}>0$. This helps to suppress
combinatorial as well as doubly misidentified backgrounds.
For the same reason the pion is required to have $\textrm{DLL}_{K-\pi}<5$.
The contamination of  \LbLcp, $\Lcbar \to \antiproton K^+ \pi^-$ is reduced by applying
a requirement of $\textrm{DLL}_{K-p}>0$ to the candidates that, when
reconstructed under the $\Lcbar \to \antiproton K^+ \pi^-$ mass hypothesis,
lie within $\pm21$~MeV$/c^2$ of the $\Lcbar$ mass.

Because of its larger branching fraction, \BsDp is a significant background to \BsDK.
It is suppressed by demanding that
the bachelor satisfies the criterion $\textrm{DLL}_{K-\pi}>5$. Conversely, a sample
of \BsDp, free of \BsDK contamination, is obtained by requiring that the bachelor satisfies
$\textrm{DLL}_{K-\pi}<0$.
The efficiency and misidentification probabilities for the PID criterion used
to select the bachelor, \Dm, and \Dsm candidates are summarised in Table~\ref{t:pid}.

\begin{table}
  \begin{center}
  \label{t:pid}   
  \caption{PID efficiency and misidentification probabilities, separated according to the up (U) and down (D) magnet
polarities. The first two lines refer to the bachelor track selection,
         the third line is the \Dm efficiency and the fourth the \Dsm efficiency. Probabilities
         are obtained from the efficiencies in the \Dstarp calibration sample,
         binned in momentum and $p_{\rm T}$. Only bachelor tracks with momentum below
         100~GeV$/c$ are considered. The uncertainties shown are the statistical uncertainties due to
         the finite number of signal events in the PID calibration samples.}
    \vspace{2.0mm}
    \begin{tabular}{ llcccc }
    \hline
     & PID Cut & \multicolumn{2}{c}{Efficiency ($\%$)} & \multicolumn{2}{c}{Misidentification ($\%$)} \\
     &         & U & D & U & D \\
    \hline
    $K$ & $\textrm{DLL}_{K-\pi}>5$ & $83.3 \pm 0.2$ & $83.5 \pm 0.2$  & $5.3 \pm 0.1$  & $4.5 \pm 0.1$   \\
    $\pi$ & $\textrm{DLL}_{K-\pi}<0$  & $84.2 \pm 0.2$& $85.8 \pm 0.2$   & $5.3 \pm 0.1$   & $5.4 \pm 0.1$ \\
    \Dm & & $ 84.1 \pm 0.2$ & $85.7 \pm 0.2$   & - & - \\
    \Dsm &   & $ 77.6 \pm 0.2$ & $78.4 \pm 0.2$ & - & - \\
    \hline
    \end{tabular}
  \end{center}
\end{table}

\section{Mass fits}
\label{sec:Fits}

The fits to the invariant mass distributions of the \BsDp and \BsDK candidates require 
knowledge of the signal and background shapes.
The signal lineshape is taken from a fit to simulated signal events which had the full
trigger, reconstruction, and selection chain applied to them. 
Various lineshape parameterisations have been examined. The best fit to the simulated event 
distributions is obtained with the sum of two Crystal Ball functions~\cite{Skwarnicki:1986} with a common peak position
and width, and opposite side power-law tails.
Mass shifts in the signal peaks relative to world average values~\cite{PDG}, arising from
an imperfect detector alignment \cite{detAligm}, are observed in the data and are accounted for.
A constraint on the \Dsm meson mass is used to improve the \Bs mass resolution.
Three kinds of backgrounds need to be considered: fully reconstructed (misidentified) backgrounds,
partially reconstructed backgrounds with or without 
misidentification (\eg \BsDstarK or \BsDrho), and combinatorial backgrounds.

The three most important fully reconstructed
backgrounds are \BdDsK and \BsDp for \BsDK,
and \BdDp for \BsDp.
The mass distribution of the \BdDp events does not suffer
from fully reconstructed backgrounds.
In the case of the \BdDsK decay, which is fully reconstructed under its own mass hypothesis,
the signal shape is fixed to be the same as for \BsDK and the peak position is varied.
The shapes of the misidentified backgrounds \BdDp and \BsDp are taken from data
using a reweighting procedure. First, a clean signal sample of \BdDp and \BsDp
decays is obtained by applying the PID selection for the bachelor track given in Sect.~\ref{sec:PID}. The invariant mass 
of these decays under the wrong mass hypothesis (\BsDp or \BsDK) 
depends on the momentum of the misidentified particle. This momentum distribution
must therefore be reweighted by taking into account the momentum dependence of the misidentifaction rate.
This dependence is obtained using a dedicated calibration sample of prompt \Dstarp decays.
The mass distributions under the wrong mass hypothesis are then reweighted
using this momentum distribution to obtain the \BdDp and \BsDp mass shapes under the \BsDp and \BsDK
mass hypotheses, respectively.

For partially reconstructed backgrounds, the probability density functions (PDFs) of the invariant mass distributions
are taken from samples of
simulated events generated in specific exclusive modes and are corrected for mass shifts, 
momentum spectra, and PID efficiencies in data.
The use of simulated events is justified by the observed good agreement between data
and simulation.

The combinatorial background in the \BsDp and \BdDp fits is modelled by an
exponential function where the exponent is allowed to vary in the fit. 
The resulting shape and normalisation of the combinatorial backgrounds are in agreement
within one standard deviation with the distribution of a wrong-sign control sample
(where the $D^{-}_{s}$ and the bachelor track have the same charges).
The shape of the combinatorial background in the \BsDK fit cannot be left free because of the partially 
reconstructed backgrounds which dominate
in the mass region below the signal peak. In this case, therefore, the combinatorial slope is
fixed to be flat, as measured from the wrong sign events.

In the \BsDK fit, an additional complication arises due to backgrounds from \LbDsp and \LbDsstp, which fall
in the signal region when misreconstructed.
To avoid a loss of \BsDK signal, no requirement is made on the $\textrm{DLL}_{K-p}$ of the bachelor 
particle. Instead, the \LbDsp mass shape is obtained 
from simulated \LbDsp decays, which are reweighted in momentum
using the efficiency of the $\textrm{DLL}_{K-\pi}> 5$ requirement on protons.
The \LbDsstp mass shape is obtained by shifting the \LbDsp mass shape downwards by 200~MeV$/c^2$.
The branching fractions of \LbDsp and \LbDsstp are assumed to be
equal, motivated by the fact that the decays $B^0\to D^-D_s^+$ and
$B^0\to D^-D_s^{*+}$ (dominated by similar tree topologies) have 
almost equal branching fractions. 
Therefore the overall mass shape is formed by summing the \LbDsp and \LbDsstp
shapes with equal weight.

The signal yields are obtained from unbinned extended maximum likelihood fits
to the data. In order to achieve the highest sensitivity, the sample is 
separated according to the two magnet polarities, 
allowing for possible differences in PID
performance and in running conditions.
A simultaneous fit to the two magnet polarities is performed for each decay,
with the peak position and width of each signal, as well as the combinatorial background shape,
shared between the two.

\begin{figure}[t]
  \centering
  \includegraphics[width=.75\textwidth]{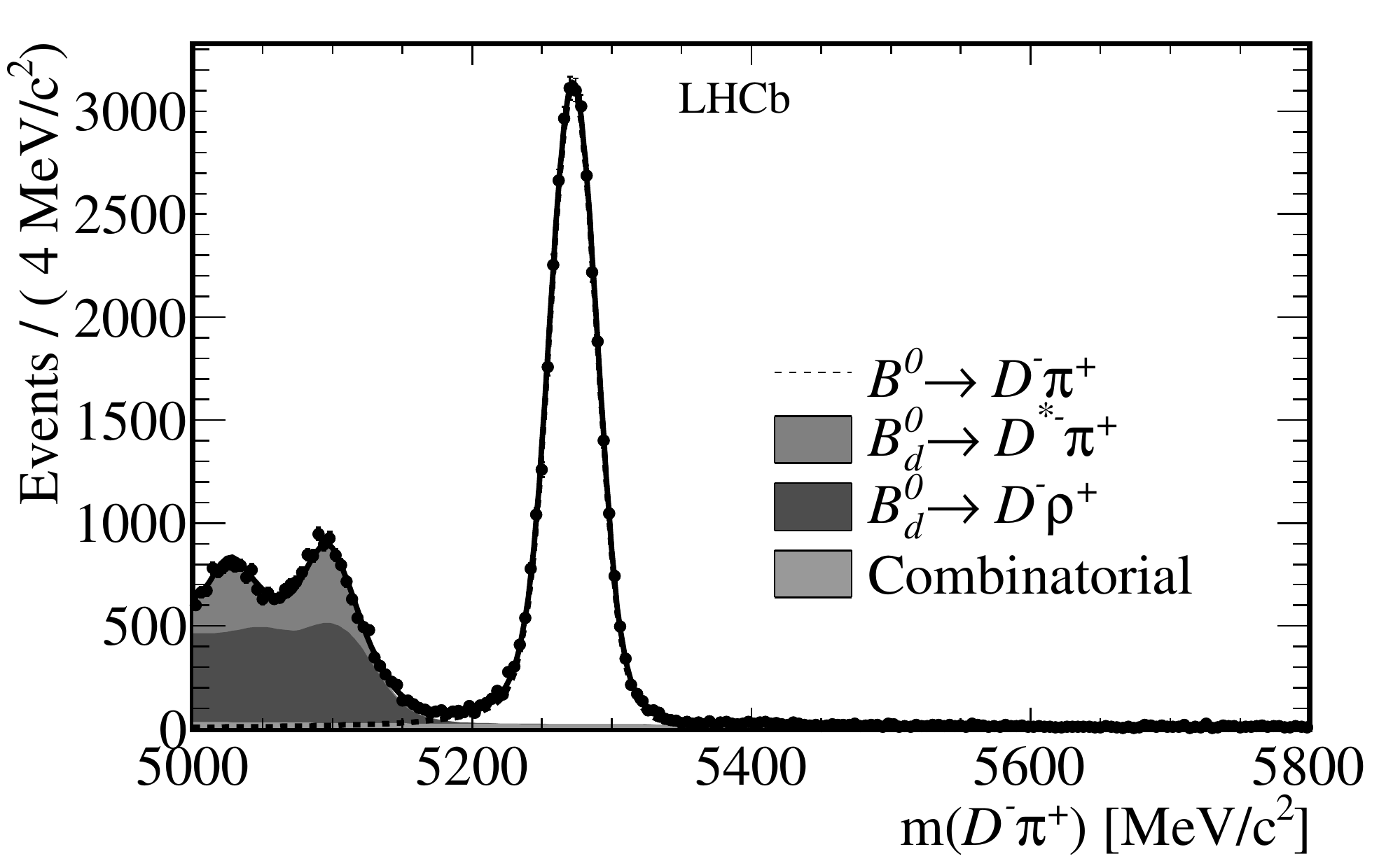}
  \includegraphics[width=.75\textwidth]{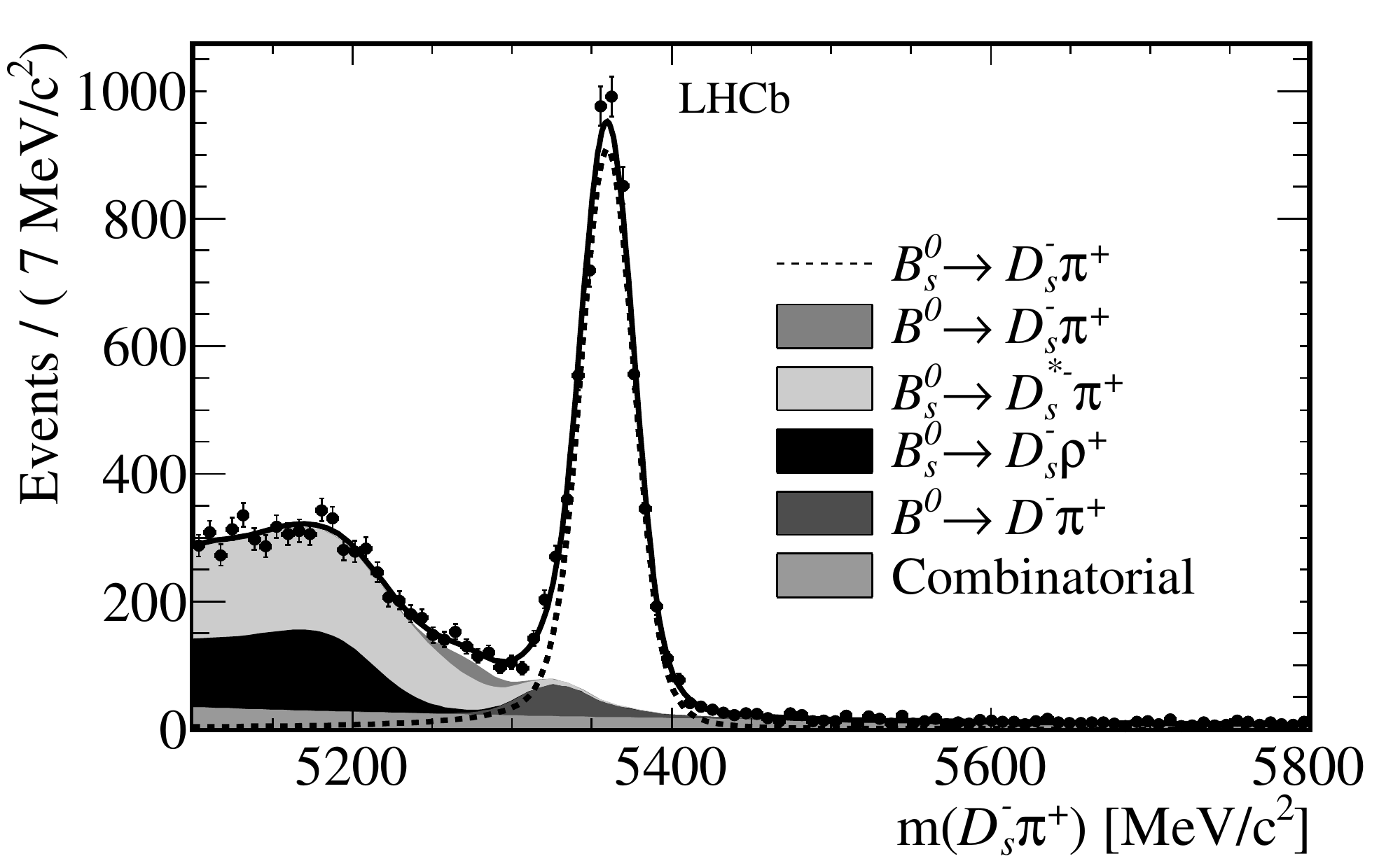}
  \caption{Mass distribution of the \BdDp candidates (top) and \BsDp candidates (bottom). The stacked background shapes follow the same top-to-bottom order in the legend and the plot. For illustration purposes the plot includes events from both magnet polarities, but they are fitted separately as described in the text.}
  \label{fig:BsDsPi_fit}
\end{figure}

The fit under the \BsDp hypothesis requires a description of
the \BdDp background.
A fit to the \BdDp spectrum is first performed to determine the yield of
signal \BdDp events, shown in Fig.~\ref{fig:BsDsPi_fit}. 
The expected \BdDp contribution under the \BsDp hypothesis
is subsequently constrained with a $10\%$ uncertainty 
to account for uncertainties on the PID efficiencies.
The fits to the \BsDp candidates are shown in Fig.~\ref{fig:BsDsPi_fit} and
the fit results for both decay modes are summarised in Table~\ref{t:allfitresults}.
The peak position of the signal shape is varied,
as are the yields of the different partially reconstructed backgrounds
(except \BdDp) and the shape of the combinatorial background.
The width of the signal is fixed to the values found in the \BdDp fit
($17.2$~MeV$/c^2$), scaled by the ratio of widths observed in simulated events
between \BdDp and \BsDp decays ($0.987$). The accuracy of these fixed parameters
is evaluated using ensembles of simulated experiments described in Sect.~\ref{sec:Systematics}.
The yield of \BdDsp is fixed to be 2.9\% of the \BsDp signal 
yield, based on the world average branching fraction of \BdDsp of
$(2.16 \pm 0.26) \times 10^{-5}$,
the value of \fsfdt given in \cite{Aaij:2011jp}, and the value of the 
branching fraction computed in this paper. The shape used to fit this component is the sum of two Crystal Ball functions
obtained from the \BsDp sample with the peak position fixed to the value obtained
with the fit of the \BdDp data sample and the width fixed to the width of the \BsDp peak.

The \LbLcp background is negligible in this fit owing to the
effectiveness of the veto procedure described earlier.
Nevertheless, a \LbLcp component, whose yield is allowed to vary, is included in the fit
(with the mass shape obtained using the reweighting procedure on simulated
events described previously) and results in a negligible contribution, as expected.

\begin{table}[t]
  \begin{center}
  \label{t:allfitresults}
  \caption{Results of the mass fits to the \BdDp, \BsDp, and \BsDK candidates separated according to the up (U) and down (D) magnet
polarities. In the \BsDK case,
  the number quoted for \BsDp also includes a small
  number of \BdDp events which have the same mass shape (20 events from the expected misidentification).
  See Table 3 for the constrained values used in the \BsDK decay fit for the partially reconstructed backgrounds and the \BDK decay channel.
}
  \vspace{2.0mm}
    \begin{tabular}{lcccccc}
      \hline
        Channel   & \multicolumn{2}{c}{\BdDp} & \multicolumn{2}{c}{\BsDp} & \multicolumn{2}{c}{\BsDK} \\
                  & U & D & U & D & U & D \\
	\hline
	$N_{\textrm{Signal}}$   & $16304\pm137$ & $20150\pm152$  & $2677\pm62$ & $3369\pm69$ & $195\pm18$ & $209\pm19$ \\
	$N_{\textrm{Comb}}$  & $\;\,1922\pm123$ & $\;\,2049\pm118$ & $\;\,869\pm63$& $\;\,839\pm47$  & $149\pm25$ & $255\pm30$ \\
        $N_{\textrm{Part-Reco}}$ & $10389\pm407$ & $12938\pm441$  & $2423\pm65$ &$3218\pm69$  & - & -\\
        $N_{\BdDsK}$ & - & - & - & - & $\;\,87\pm17$ & $100\pm18$ \\
	$N_{\BsDp}$ &  - & - & - & -  & $154\pm20$ & $164\pm22$\\
        \hline
    \end{tabular}
  \end{center}
\end{table}

\begin{figure}[htb]
  \centering
  \includegraphics[width=.95\textwidth]{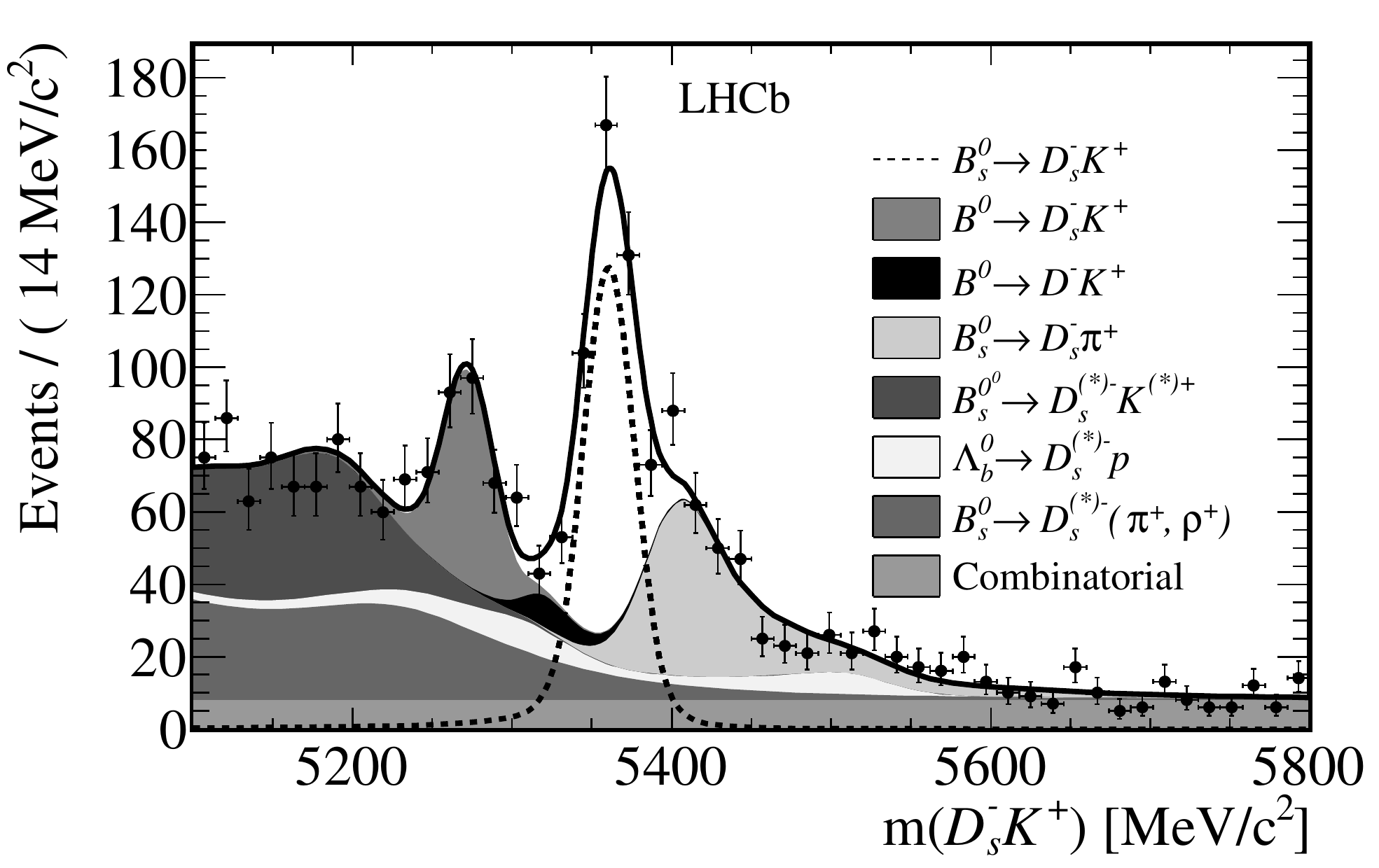}
  \caption{Mass distribution of the \BsDK candidates. The stacked background shapes follow the same top-to-bottom order in the legend and the plot. For illustration purposes the plot includes
events from both magnet polarities, but they are fitted separately as described in the text.}
  \label{fig:BsDsK_fit}
\end{figure}

The fits for the \BsDK candidates are shown in Fig.~\ref{fig:BsDsK_fit} and
the fit results are collected in Table~\ref{t:allfitresults}.
There are numerous reflections which contribute to the mass distribution.
The most important reflection is \BsDp, whose shape is taken 
from the earlier \BsDp signal fit, reweighted according to the efficiencies of the applied PID requirements.
Furthermore, the yield of the \BdDK reflection is
constrained to the values in Table~\ref{t:dskconstraints}.
In addition, there is 
potential cross-feed from partially
reconstructed modes with a misidentified pion such as \BsDrho,
as well as several small contributions from partially reconstructed backgrounds
with similar mass shapes.  
The yields of these modes, whose branching fractions are
known or can be estimated (\eg \BsDrho, \BsDkst), are constrained to the values
in Table~\ref{t:dskconstraints}, based on criteria such as relative branching fractions
and reconstruction efficiencies and PID probabilities.
An important cross-check is performed by comparing the fitted value 
of the yield of misidentified \BsDp events ($318\pm30$) 
to the yield expected from PID efficiencies ($370\pm11$) and an agreement is found.

\begin{table}
\begin{center}
  \label{t:dskconstraints} 
  \caption{Gaussian constraints on the yields of partially reconstructed and misidentified backgrounds applied in the \BsDK fit,
separated according to the up (U) and down (D) magnet polarities.}
    \vspace{2.0mm}
    \begin{tabular}{ lcc }
    \hline
    Background type   & U          & D   \\
    \hline			     		
    \BDK               &  $\,16  \pm  \,3$ & $\,17 \pm  \,3$ \\
    \BsDstarp          &  $\;\,63  \pm  21$ &  $\;\,70 \pm  23$\\ 
    \BsDstarK          &  $\;\,72  \pm  34$ &  $\;\,80 \pm  27$\\
    \BsDrho            &  $135     \pm  45$ &  $150    \pm  50$\\
    \BsDkst            &  $135     \pm  45$ &  $150    \pm  50$\\
    \BsDstrho          &  $\;\,45  \pm  15$ &  $\;\,50 \pm  17$\\
    \BsDstkst          &  $\;\,45  \pm  15$ &  $\;\,50 \pm  17$\\
    \LbDsp + \LbDsstp  &  $\;\,72  \pm  34$ &  $\;\,80 \pm  27$\\
    \hline
    \end{tabular}
  \end{center} 
\end{table}
\section{Systematic uncertainties}
\label{sec:Systematics}
The major systematic uncertainities on the measurement of the relative branching fraction of
\BsDK and \BsDp are related to the fit, PID calibration,
and trigger and offline selection efficiency corrections.
Systematic uncertainties related to the fit are evaluated by generating large sets of simulated 
experiments using the nominal fit, and then fitting them with a model where certain parameters are varied. 
To give two examples, the signal width is deliberately fixed to a value different from the width used in the generation,
or the combinatorial background slope in the \BsDK fit is fixed to the
combinatorial background slope found in the \BsDp fit.
The deviations of the peak position of the pull distributions from zero are 
then included in the systematic uncertainty.
\begin{table}[t]
  \begin{center}
  \label{t:sysbudgettotal}   
  \caption{Relative systematic uncertainities on the branching fraction ratios.  }
    \vspace{2.0mm}
    \begin{tabular}{ lccc }
    \hline 
    Source \TVA \BVA & $\frac{\BsDK}{\BsDp} (\%)$ & $\frac{\BsDp}{\BdDp}(\%)$ & $\frac{\BsDK}{\BdDp}(\%)$ \\
    \hline
    All non-PID selection & $2.0$ & $2.0$ & $3.0$ \\
    PID selection  & $1.8$ & $1.3$ & $2.2$\\    
    Fit model      & $2.4$ & $1.7$ & $2.2$\\
    Efficiency ratio & $1.5$ & $1.6$ & $1.6$ \\
    \hline
    Total & $3.9$ & $3.4$ & $4.6$\\
    \hline
    \end{tabular}
  \end{center}
\end{table}

In the case of the \BsDK fit the presence of constraints
for the partially reconstructed backgrounds must be considered. 
The generic extended likelihood function can be written as
\begin{equation}
	\mathcal{L} = e^{-N} N^{N_{\rm obs}}
		\times \prod_j G(N^j; N^j_c, \sigma_{N^j_0})
		\times \prod_{i=1}^{N_{\rm obs}} P(m_i; \vec{\lambda})\,,
\end{equation}
where the first factor is the extended Poissonian likelihood in which $N$ is the 
total number of fitted events, given by the sum of the fitted
component yields $N = \sum_k N_k$. The fitted data sample contains $N_{\rm obs}$
events. The second factor is the product of the $j$ external constraints on the yields, 
$j<k$, where $G$ stands for a Gaussian PDF, and $N_c \pm \sigma_{N_0}$ is the constraint
value. The third factor is a product over all events in the sample, $P$ is the
total PDF of the fit, $P(m_i; \vec{\lambda}) = \sum_k N_k P_k(m_i; \vec{\lambda}_k)$, and $\vec{\lambda}$
is the vector of parameters that define the mass shape and are not fixed in the fit.
 
Each simulated dataset is generated by first varing the component yield $N_k$
using a Poissonian PDF, then sampling the resulting number of events from $P_k$,
and repeating the procedure for all components. In addition, constraint values $N_c^j$
used when fitting the simulated dataset are generated by drawing from $G(N; N^j_0, \sigma_{N^j_0})$,
where $N^j_0$ is the true central value of the constraint, while in the nominal fit
to the data  $N_c^j = N^j_0$. 

The sources of systematic uncertainty considered for the fit are signal widths,
the slope of the combinatorial backgrounds, and constraints placed on specific backgrounds.
The largest deviations are due to the signal widths and the fixed slope of the combinatorial
background in the \BsDK fit.

The systematic uncertainty related to PID enters in two ways:
firstly as an uncertainty on the overall efficiencies and misidentification
probabilities, and secondly from the shape for the misidentified 
backgrounds which
relies on correct reweighting of PID efficiency versus momentum.
The absolute errors on the individual $K$ and $\pi$ efficiencies,
after reweighting of the \Dstarp calibration sample, have been determined
for the momentum spectra that are relevant for this analysis, and are found to be
0.5\% for $\textrm{DLL}_{K-\pi}< 0$ and 
0.5\% for $\textrm{DLL}_{K-\pi}> 5$.

The observed signal yields are corrected by the difference observed 
in the (non-PID) selection efficiencies of different modes as measured from simulated
events:
\begin{eqnarray*}
\epsilon(\BsDp)/\epsilon(\BdDp) &=& 1.015\;,\\
\epsilon(\BsDp)/\epsilon(\BsDK) &=& 1.061\;. 
\end{eqnarray*}
A systematic uncertainty is assigned on the ratio to account for percent level differences 
between the data and the simulation. These are dominated by the simulation of the hardware trigger.
All sources of systematic uncertainty are summarized in Table~\ref{t:sysbudgettotal}.
\section{Determination of the branching fractions}
\label{sec:Results}

The \BsDK branching fraction relative to \BsDp is obtained by correcting the raw
signal yields for PID and selection efficiency differences

\begin{equation}
\frac{\BR\left(\BsDK\right)}{\BR\left(\BsDp\right)} =
   \frac{N_{\BsDK}}{N_{\BsDp}}
   \frac{\epsilon^{\textrm{PID}}_{\BsDp}}{\epsilon^{\textrm{PID}}_{\BsDK}}
   \frac{\epsilon^{\textrm{Sel}}_{\BsDp}}{\epsilon^{\textrm{Sel}}_{\BsDK}}\;,
\end{equation}
where $\epsilon_{X}$ is the efficiency to reconstruct decay mode $X$ and $N_{X}$ is 
the number of observed events in this decay mode. The PID efficiencies are given in Table~\ref{t:pid}, and the ratio of the two selection efficiencies is $0.943\pm0.013$.

The ratio of the branching fractions of \BsDK relative to \BsDp is 
determined separately for the down ($0.0601\pm 0.0056$) and up ($0.0694\pm 0.0066$) magnet polarities
and the two results are in good agreement. The quoted errors are purely statistical.
The combined result is
\begin{equation*}
\label{eq:Bsratio}
\frac{\BR\left(\BsDK\right)}{\BR\left(\BsDp\right)} = 0.0646
\pm 0.0043 \pm 0.0025 \;,
\end{equation*}
where the first uncertainty is statistical and the second is the total systematic
uncertainty from Table~\ref{t:sysbudgettotal}. 

The relative yields of \BsDp and \BdDp are used to extract the branching fraction of
\BsDp from the following relation 
\begin{equation}
\BR(\BsDp)= \BR\left(\BdDp\right)
\frac{\epsilon_{\BdDp}}{\epsilon_{\BsDp}}
\frac{N_{\BsDp}\BR\left(D^- \to K^+ \pi^-\pi^-\right)}{\fsfd N_{\BdDp}
\BR\left(D^-_s \to K^- K^+ \pi^-\right)}\;,
\end{equation}
using the recent \fsfdt measurement from semileptonic decays~\cite{Aaij:2011jp} 
\begin{equation*}
\fsfd = 0.268 \pm 0.008^{+0.022}_{-0.020}\;,
\end{equation*}
where the first uncertainty is statistical and the second systematic. Only the semileptonic result is used
since the hadronic determination of \fsfdt relies on theoretical assumptions about the ratio of the branching
fractions of the \BsDp and \BdDp decays.
In addition, the following world average values~\cite{PDG} for the $B$ and $D$ branching fractions are used
\begin{eqnarray*}
\BR(\BdDp)  &=& \left(2.68\pm0.13\right)\times 10^{-3}\;,\\
\BR(\Dm \to K^+ \pi^- \pi^-) &=&  \left(9.13\pm0.19\right)\times 10^{-2}\;,\\
\BR(\Dsm \to K^+ K^- \pi^-) &=&  \left(5.49\pm0.27\right)\times10^{-2}\;,
\end{eqnarray*}
leading to 
\begin{eqnarray*}
\BR(\BsDp) &=& (2.95 \pm 0.05 \pm 0.17^{+0.18}_{-0.22})\times 10^{-3}\;,\\
\BR(\BsDK) &=& (1.90 \pm 0.12 \pm 0.13^{+0.12}_{-0.14})\times 10^{-4}\;,
\end{eqnarray*}
where the first uncertainty is statistical, the second is the experimental systematics
(as listed in Table~\ref{t:sysbudgettotal}) plus the uncertainty arising from the \BdDp branching
fraction, and the third is the uncertainty (statistical and systematic) from the
semileptonic \fsfdt measurement. Both measurements are significantly more precise than the existing world averages~\cite{PDG}.

\section*{Acknowledgments}
\noindent We express our gratitude to our colleagues in the CERN accelerator
departments for the excellent performance of the LHC. We thank the 
technical and administrative staff at CERN and at the LHCb institutes,
and acknowledge support from the National Agencies: CAPES, CNPq,
FAPERJ and FINEP (Brazil); CERN; NSFC (China); CNRS/IN2P3 (France);
BMBF, DFG, HGF and MPG (Germany); SFI (Ireland); INFN (Italy); FOM and 
NWO (The Netherlands); SCSR (Poland); ANCS (Romania); MinES of Russia and 
Rosatom (Russia); MICINN, XuntaGal and GENCAT (Spain); SNSF and SER 
(Switzerland); NAS Ukraine (Ukraine); STFC (United Kingdom); NSF 
(USA). We also acknowledge the support received from the ERC under FP7 
and the Region Auvergne.

\addcontentsline{toc}{section}{References}
\bibliographystyle{LHCb}
\bibliography{main}

\end{document}